\documentclass[aps,twocolumn,reprint,amsmath,amssymb,superscriptaddress,floatfix]{revtex4-1}

\usepackage[breaklinks=true,colorlinks,citecolor=blue,linkcolor=blue,urlcolor=blue]{hyperref}
\usepackage{float}
\usepackage{epsfig,mathrsfs,color,latexsym,subfigure,marginnote,graphicx,verbatim,relsize,mathrsfs,color,array,amsmath,amsfonts,amssymb,graphicx}
 \usepackage{multirow}
 \usepackage{makecell}

\def\be{\begin{equation}}
\def\ee{\end{equation}}

\def\bea{\begin{eqnarray}}
\def\eea{\end{eqnarray}}

\newcommand{\rsp}[1]{\textcolor{black}{#1}}

\begin{document}
\title{Tunable spin polarization and electronic structure of bottom-up synthesized MoSi$_2$N$_4$ materials} 

\author{Rajibul Islam}
\thanks{These authors contributed equally to this work}
\affiliation{International Research Centre MagTop, Institute of Physics, Polish Academy of Sciences, Aleja Lotnikow 32/46, PL-02668 Warsaw, Poland}

\author{Barun Ghosh}
\thanks{These authors contributed equally to this work}
\affiliation{Department of Physics, Indian Institute of Technology, Kanpur 208016, India}
\affiliation{Department of Physics, Northeastern University, Boston, Massachusetts 02115, USA}

\author{Carmine Autieri}
\affiliation{International Research Centre MagTop, Institute of Physics, Polish Academy of Sciences, Aleja Lotnikow 32/46, PL-02668 Warsaw, Poland}

\author{Sugata Chowdhury}
\affiliation {Department of Physics and Astronomy, Howard University, Washington, D.C. 20059, USA}
\affiliation {IBM-HBCU Quantum Center, Howard University, Washington, D.C. 20059, USA}

\author{Arun Bansil}
\affiliation{Department of Physics, Northeastern University, Boston, Massachusetts 02115, USA}

\author{Amit Agarwal}
\affiliation{Department of Physics, Indian Institute of Technology, Kanpur 208016, India}

\author{Bahadur Singh}
\email{bahadur.singh@tifr.res.in}
\affiliation{Department of Condensed Matter Physics and Materials Science, Tata Institute of Fundamental Research, Mumbai 400005, India}

\begin{abstract}
Manipulation of spin-polarized electronic states of two-dimensional (2D) materials under ambient conditions is necessary for developing new quantum devices with small physical dimensions. Here, we explore spin-dependent electronic structures of ultra-thin films of recently introduced 2D synthetic materials MSi$_2$Z$_4$ (M = Mo or W and Z = N or As) using first-principles modeling. Stacking of MSi$_2$Z$_4$ monolayers is found to generate dynamically stable bilayer and bulk materials with thickness-dependent properties. When spin-orbit coupling (SOC) is included in the computations, MSi$_2$N$_4$ monolayers display indirect bandgaps and large spin-split states at the $K$ and $K'$ symmetry points at the corners of the Brillouin zone with nearly 100\% spin polarization. The spins are locked in opposite directions along an out-of-the-plane direction at $K$ and $K'$, leading to spin-valley coupling effects. As expected, spin polarization is absent in the pristine bilayers due to the presence of inversion symmetry, but it can be induced via an external out-of-plane electric field much like the case of Mo(W)S$_2$ bilayers. A transition from an indirect to a direct bandgap can be driven by replacing N by As in MSi$_2$(N, As)$_4$ monolayers. Our study indicates that the MSi$_2$Z$_4$ materials can provide a viable alternative to the MoS$_2$ class of 2D materials for valleytronics and optoelectronics applications. 
\end{abstract}

\maketitle

{\it Introduction.}
Since the isolation of two-dimensional (2D) graphene from its parent graphite in 2004~\cite{novoselov2004electric,novoselov2005two,zhang2005experimental}, a variety of atomically thin materials have been  exfoliated from bulk layered compounds with electronic states that encompass insulators to semiconductors to semimetals/metals. Prominent examples include hexagonal boron nitride~\cite{lin2010soluble}, 2D transition-metal dichalcogenides (TMDs)~\cite{PhysRevLett.105.136805,wang2012electronics,chhowalla2013chemistry,huang2013metal,C4CS00182F,bilayer_spin,Chang2014,CSuggest_1}, phosphorene~\cite{li2014black,liu2015semiconducting}, and MXenes~\cite{naguib201425th}, among other materials~\cite{Review_NovoNeto}. These 2D materials offer exciting opportunities for exploring novel electronic, excitonic, correlated, and topological states under 2D charge confinement for spintronics, valleytronics, and optoelectronics applications and developing materials platforms for high-density devices with minimal physical dimensions. Stacking, twisting, and straining of such 2D layers to form moire superlattices and heterostructures brings unprecedented possibilities for tailoring properties~\cite{novoselov2012two,heine2015transition,gong2014vertical,wang2017graphene,tblg,Review_NovoNeto,CDW_BS2,hetrocryst_Bansil}. A common approach for obtaining 2D materials is exfoliation from appropriate 3D layered materials using a top-to-bottom approach. Finding new 2D materials without parental analogs would provide a new paradigm for engineering states with diverse functionalities and offer new pathways for designing synthetic materials with desirable properties~\cite{novoselov2012two,heine2015transition,gong2014vertical,wang2017graphene,tblg,Review_NovoNeto,CDW_BS2,hetrocryst_Bansil,Review_NovoNeto}. 

Among the methods of growing materials in a bottom-up approach is the use of a substrate with strong adatom adhesion. This method has shown success in synthesizing atomically thin films such as silicene~\cite{PhysRevLett.108.245501}, germanene~\cite{germanene}, bismuthene~\cite{Cheng2014}, and borophene~\cite{PhysRevLett.118.096401}. The stability and morphology of such materials are, however, strongly dependent on growth conditions due to the presence of dangling bonds of adatoms that either reorganize to generate complicated surface morphologies or get oxidized when exposed to air~\cite{silicene_temp}. An alternate route proposed recently involves passivation of the high-energy surfaces of materials with elements that can generate synthetic layered 2D materials~\cite{hong2020chemical,MoSiN_Novo}.  By passivating non-layered molybdenum nitride with elemental silicon during chemical vapor deposition growth, large area (15 mm$\times$ 15 mm) layered 2D MoSi$_2$N$_4$ materials were synthesized. Importantly, MoSi$_2$N$_4$ shows remarkable properties such as stability under ambient conditions, a semiconducting behavior, and high mobility of ~270/1200 cm$^2$V$^{-1}$s$^{-1}$, which is better than that of the widely used MoS$_2$ class of 2D materials~\cite{hong2020chemical,MoSiN_Novo,bertolazzi2011stretching,cai2014polarity}. MoSi$_2$N$_4$ and its derivative monolayers host gapped states in a pair of valleys located at the corners of the hexagonal Brillouin zone (BZ)~\cite{li2020valley,ai2021theoretical,Yang_valley}. Due to the breaking of the spatial inversion symmetry, the spin states in these monolayers become separated in energy and give rise to unique spin-valley couplings in the vicinity of the Fermi level and valley-contrasting Berry curvatures and orbital magnetic moments, which could potentially enable wide-ranging valleytronics and optoelectronics applications~\cite{PhysRevLett.97.186404, PhysRevLett.99.236809, PhysRevLett.106.136806, PhysRevLett.108.196802, PhysRevB.88.115140, xu2014spin, Carmine}. \rsp{Despite the excellent stability of synthetic MoSi$_2$N$_4$ monolayers under ambient conditions, it is not clear how their properties evolve in the multilayer and bulk of these bottom-up grown 2D van der Waals (vdW) materials.} 

Motivated by the new opportunities offered by a bottom-up approach, here we report layer-dependent stability and valleytronic properties of MSi$_2$Z$_4$ (M = Mo or W, and Z = N or As) materials. Using density-functional-theory based first-principles modeling, we show that the MoSi$_2$N$_4$ materials are dynamically stable up to the bulk limit. The monolayers are found to exhibit large spin-split states at the BZ corners $K$ and $K'$ with nearly 100\% spin-polarization, similar to the MoS$_2$ materials class. As expected, the spin-splitting is zero in the bilayer films as the inversion symmetry is restored. However, spin-splitting can be switched on and manipulated in the bilayers via an out-of-plane electric field. An indirect to direct bandgap transition in MSi$_2$Z$_4$ is driven by the replacement of N by As. In addition to highlighting the unique thickness-dependent properties of MSi$_2$Z$_4$, our study demonstrates the value of a bottom-up approach for synthesizing viable 3D bulk materials based on synthetic 2D vdW materials. 

{\it Methods.} 
Electronic structure calculations were performed within the density functional theory (DFT) framework using the Vienna ab-initio simulation package (VASP)~\cite{vasp1,vasp2}. The projector augmented wave (PAW) pseudopotentials were used with generalized-gradient approximation (GGA)~\cite{perdew1996generalized} for treating exchange-correlation effects. A plane-wave cutoff of 500 eV was used in all calculations. Surface BZ integrations were performed using a $10\times10\times1$ Monkhorst-pack $k-$grid. Effects of spin-orbit coupling (SOC) were included self-consistently. The structural parameters were optimized until the residual forces on each atom became less than 10$^{-4}$ eV/\AA, and these optimized parameters were used in the calculations. An energy tolerance of 10$^{-8}$ eV was used. The thin-film calculations were performed using a slab geometry with a vacuum layer of 20 {\AA} to eliminate spurious interactions between the periodically repeated 2D layers. Phonon dispersion curves were obtained within the density functional perturbation theory (DFPT) framework using PHONOPY code~\cite{togo2015first} with a $4 \times 4\times 1$ supercell. 
\rsp{The robustness of our GGA-based results was assessed using the optPBE-vdW correlation functional~\cite{vdw_optPBE_1,vdw_optPBE_2,vdw_1,vdw_2,vdw_3} as well as the more advanced HSE hybrid-functional~\cite{hse}, see Supplemental Material (SM)~\cite{SPM} for details. } 
PyProcar~\cite{herath2020pyprocar} and Pymatgen~\cite{ong2013python} packages were used for band structure illustrations.

\begin{figure}[t!] % Crystal structure and phonons of MoSi$_2$N$_4$
\centering    
\includegraphics[width=0.48\textwidth,]{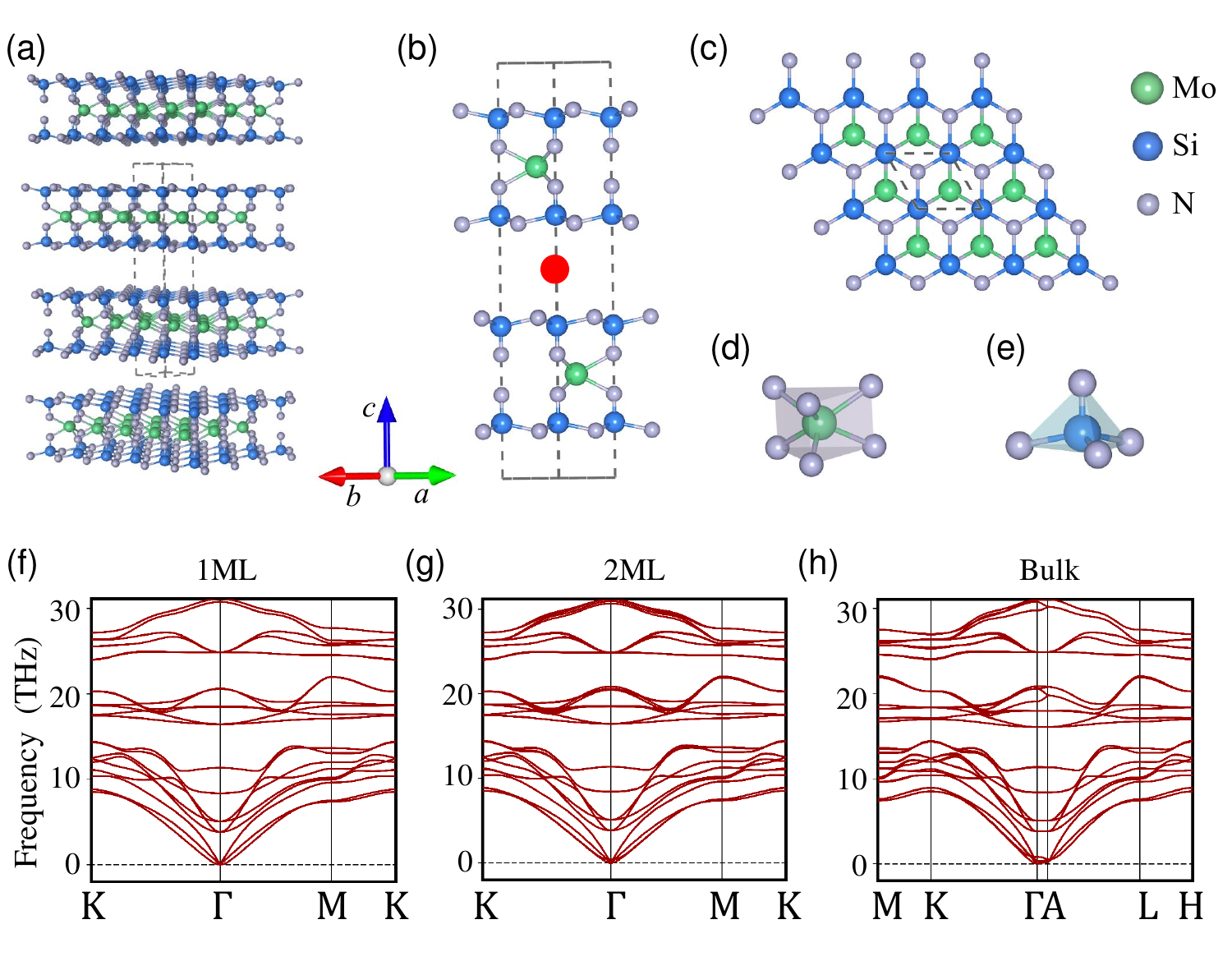}
\caption{Atomic arrangement of (a) four and (b) two layers of MoSi$_2$N$_4$ with AB stacking. The dashed box identifies the bulk unit cell of 2H-MoSi$_2$N$_4$. The red dot in the middle of the van der Waals gap in (b) marks the spatial center of inversion, which is absent in the monolayer.  (c) Top view of monolayer MoSi$_2$N$_4$. (d) Mo-N trigonal and (e) Si-N tetrahedral local coordination structures in MoSi$_2$N$_4$ monolayers. The calculated phonon dispersion of (f) monolayer (1ML), (g) bilayer (2ML), and (h) bulk MoSi$_2$N$_4$. }
\label{fig:CS}
\end{figure}

{\it Crystal structure and dynamical stability of MoSi$_2$N$_4$.} 
Monolayer MoSi$_2$N$_4$ crystallizes in the hexagonal lattice with space group $D^1_{3h}$ ($P\bar{6}m2$, No.~187). It involves strongly-bonded, seven-layer stacking in the order N-Si-N-Mo-N-Si-N that can be viewed as a sandwich involving an MoN$_2$ layer and two Si-N bilayers [Fig. \ref{fig:CS}(a)-(e)]. This structure preserves trigonal $C_{3v}$ and $M_z (z \rightarrow -z)$ mirror-plane symmetries but breaks the inversion symmetry. The monolayers can be stacked in the -A-B-A- order to realize a 2H bilayer structure similar to that of MoS$_2$. Unlike the monolayer, bilayer MoSi$_2$N$_4$ realizes the higher-symmetry group $D^4_{6h}$ ($P6_3/mmc$, No.~194)~\cite{hong2020chemical,zhong2020strain}, restoring the spatial center of inversion, which is marked by the red dot in Fig. \ref{fig:CS}(b). The equilibrium interlayer distance ($d_0$) between the Mo$_1$ and Mo$_2$ sublayers in the bilayer is found to be 10.65 \AA. Notably, the 2H-bilayer structure can be repeated to realize the bulk MoSi$_2$N$_4$ materials like the transition metal dichalcogenides. The optimized structural parameters and Wyckoff positions for bulk MSi$_2$Z$_4$ are listed in Table \ref{T1:bulk}.

In order to showcase the stability of the monolayer and multilayer MoSi$_2$N$_4$ films, we present the associated phonon dispersions in Figs. \ref{fig:CS}(f)-(h). The absence of imaginary phonon frequencies in the entire hexagonal BZ confirms the dynamical stability of these structures. Notably, the bulk phonon spectrum also lacks imaginary phonon frequencies. 
\rsp{Our computations in which van der Waals interactions beyond the GGA are included yield similar results and affirm the robustness of our conclusions concerning the stability in all cases~\cite{SPM}.}  
We thus infer that stable 3D bulk of MoSi$_2$N$_4$ can be realized experimentally~\footnote{\rsp{Notably, the thickness-dependent stability of 2D MoSi$_2$Z$_4$ materials indicates that multilayers and bulk of these materials should be possible to realize experimentally. Our results here should be contrasted with the thickness-dependent stability studies in the literature on various 2D materials synthesized using a top-to-bottom approach from a stable bulk. It is not clear a priori that stability in top-to-bottom synthesis ensures a similar stability in the bottom-up synthesis.}}.

\begin{table*}[t!]
\caption{Calculated lattice parameters for 2H-bulk MoSi$_2$N$_4$,  MoSi$_2$As$_4$, WSi$_2$N$_4$, and WSi$_2$As$_4$ 
\rsp{using the GGA and optPBE-VDW density functionals}.
 $a$ and $c$ are the hexagonal lattice constants and $u_{Si}$ , $u_{N/As}$, and $v_{N/As}$  are the internal parameter associated with Wyckoff positions $4e~(0, 0, u_{Si})$, $4f~(\frac{1}{3},\frac{2}{3}, u_{N/As})$, and $4e~(0, 0, v_{N/As})$, respectively. The subscripts identify the atoms.}

\begin{tabular}{c c c c c c c c}
\hline \hline 

    &   & a (\r{A}) 	& c (\r{A}) & $u_{Si}$    & $u_{N/As}$ 	& $v_{N/As}$ & E$_g$ (eV)\\
\hline
\multirow{2}{*} {MoSi$_2$N$_4$ }  & GGA 	& 2.910        	& 21.311	& 0.1095	  & 0.1915 	& 0.0859 		&1.655	\\  
							& vdW	& 2.932        	& 20.772	& 0.1045	  & 0.1889 	& 0.0804 		&1.665	\\  
							
\multirow{2}{*} {MoSi$_2$As$_4$ } & GGA	& 3.622   		& 27.617    & 0.1106     & 0.1960        	& 0.0703     	& 0.508	\\
  							& vdW	& 3.681   		& 27.408       	& 0.1079         	  & 0.1950       	& 0.0670     		& 0.447	\\

\multirow{2}{*} {WSi$_2$N$_4$ }	& GGA  	& 2.914  		& 21.439       	& 0.1099         	  & 0.1914        	& 0.0865   		& 1.970	\\
  							& vdW & 2.935  		& 20.763       	& 0.1043         	  & 0.1888        	& 0.0805   		& 1.985	\\

\multirow{2}{*} {WSi$_2$As$_4$ } 	&  GGA & 3.628  		& 27.940      	& 0.1121            & 0.1967       		 & 0.0723   		& 0.207	\\
							& vdW & 3.685  		& 27.397      	& 0.1079            & 0.1952       		 & 0.0672   		& 0.208	\\

\hline  \hline
\end{tabular}\label{T1:bulk}
\end{table*}

\begin{figure}[!t] % Monolayer band structure with spin-polarization
\centering
\includegraphics[width=1.04\linewidth]{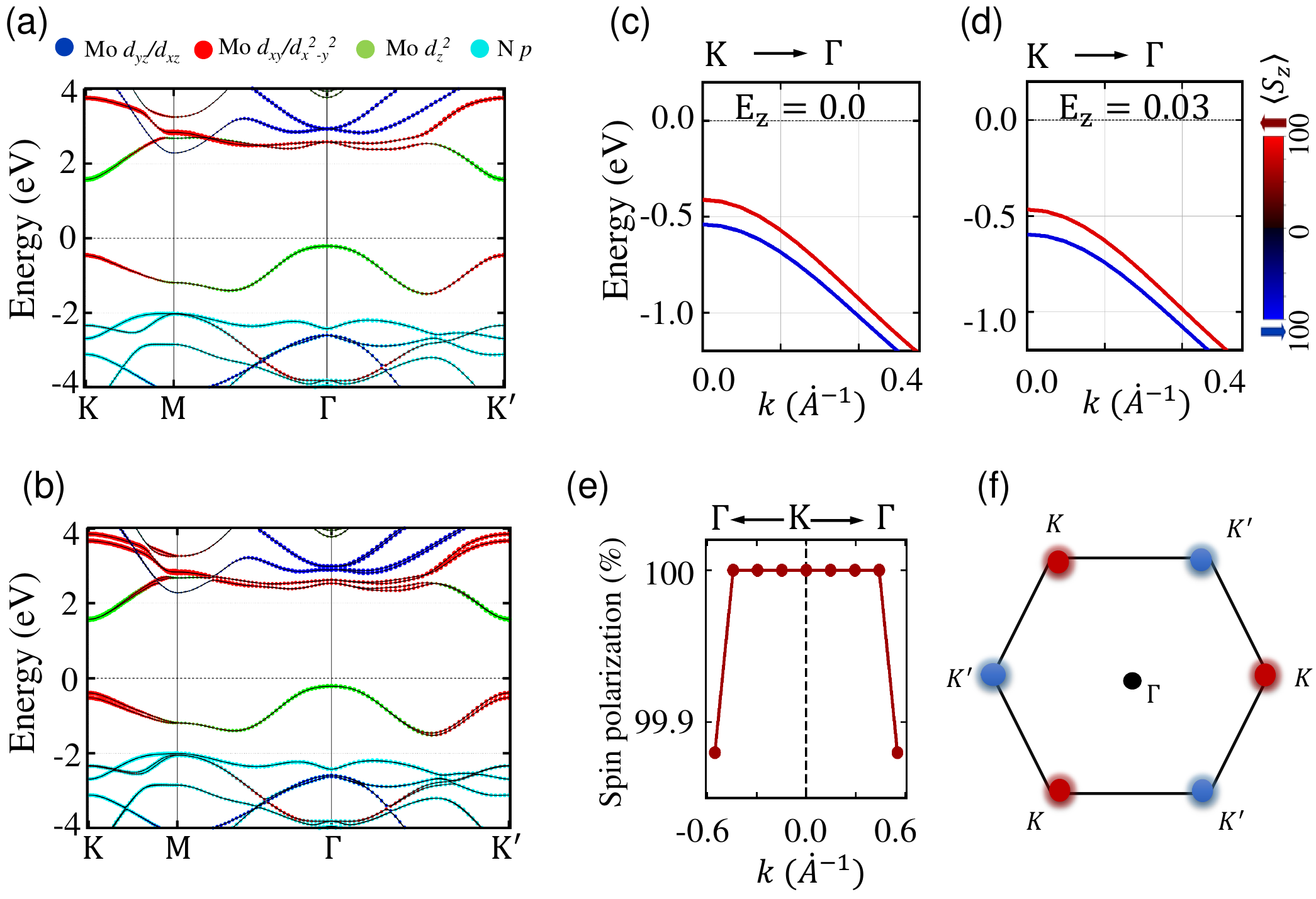}
\caption{Orbitally-resolved band structure of monolayer MoSi$_2$N$_4$ (a) without and (b) with spin-orbit coupling (SOC). Spin-resolved bands around $K$ along the $\Gamma-K$ direction for (c) $E_z$ = 0 eV/\AA~ and (d) $E_z$ = 0.03 eV/\AA~with SOC. The color bar in (d) denotes the degree (in percent) of spin-polarization. (e) Spin-polarization decay profile of the states at the top of the valence band around the $K$ point. Large spin polarization ($> 99.9\%$) persists over a wide momentum range along the $\Gamma-K$ direction. (f) Schematic representation of spin-valley locking in monolayer MoSi$_2$N$_4$. Red (blue) color represents spin pointing out of (into) the plane.
}\label{fig:MonoSpinP}
\end{figure}

{\it Spin-resolved electronic structure of monolayer MoSi$_2$N$_4$.}
The orbitally-resolved band structure of monolayer MoSi$_2$N$_4$ without SOC is presented in Fig~\ref{fig:MonoSpinP}(a). An indirect band gap of $1.778$ eV is obtained between the valence band maximum (VBM) and conduction band minimum (CBM), which are located at the $\Gamma$ and $K/K'$ points, respectively. The energy difference, $\Delta_{\Gamma K}$, between the top of the valence bands at the $\Gamma$ and $K/K'$ points is $322$ meV, and it can be tuned by strain to realize a direct band gap at the $K/K'$ point~\cite{li2020valley}.  The Bloch wave functions at the VBM and CBM edges are composed of $d_{z^2}$ states of the Mo atoms. All states remain twofold spin degenerate without the SOC as seen in Fig.~\ref{fig:MonoSpinP}(a). When SOC is included, the top of the valence bands displays a large spin-splitting of 129 meV at $K$ due to the broken spatial inversion symmetry. [Since $K$ is not a time-reversal invariant momentum (TRIM) point, the spin-split states at $K$ are not twofold degenerate.] In contrast, the bands at the $\Gamma$ and $M$ points remain twofold spin degenerate as they are TRIM points [see Figs.~\ref{fig:MonoSpinP}(b) and (c)]. The indirect nature of the monolayer band gap, however, remains preserved with a value of 
\rsp{$1.775$ eV (2.342 eV) with GGA (HSE)}. 

Our analysis reveals that the two spin-split states at $K$ have nearly 100\% out-of-plane ($S_z$) spin-polarization. This can be attributed to the presence of the horizontal mirror plane $M_z$ in monolayer MoSi$_2$N$_4$ that ensures that the $S_x$ and $S_y$ components of spin are zero. The spin-split states at $K$ and $K'$ are oppositely polarized since they form a Kramers pair obeying the time-reversal symmetry constraint $E(\vec{k},\uparrow)$ = $E(-\vec{k},\downarrow)$. Figure~\ref{fig:MonoSpinP}(d) shows the evolution of the degree of spin-polarization of states at the top of the valence band as we go away from the $K$ point. Spin polarization decreases slightly to 99.9\% for the change a momentum $\Delta k=$0.553 \AA$^{-1}$ ($\sim$38\% of the $\Gamma-K$ distance), demonstrating its robustness. The spin texture of the state at the top of the valence band in the hexagonal BZ is shown schematically in Fig. ~\ref{fig:MonoSpinP}(f). The preceding spin behavior is indicative of spin-valley locking in MoSi$_2$N$_4$ monolayers, which is similar to that observed previously in the TMDs~\cite{Chang2014}.

We emphasize that the Zeeman-type out-of-plane spin polarization in the vicinity of $K$ points in the MoSi$_2$N$_4$ monolayer is tied to the crystal structure of the film, and therefore, it cannot be destroyed or manipulated with an out-of-plane electric field $E_{z}$. We have verified this property by calculating the spin-resolved band structure in the presence of an external electric field applied perpendicular to the monolayer.  Figure~\ref{fig:MonoSpinP}(d) shows the results for $E_z= 0.03$ eV/{\AA}. Both the spin-splitting and spin-polarization features are seen to be retained.

\begin{figure}[!t] % bilayer band structure with spin-polarization
\centering    
\includegraphics[width=\linewidth]{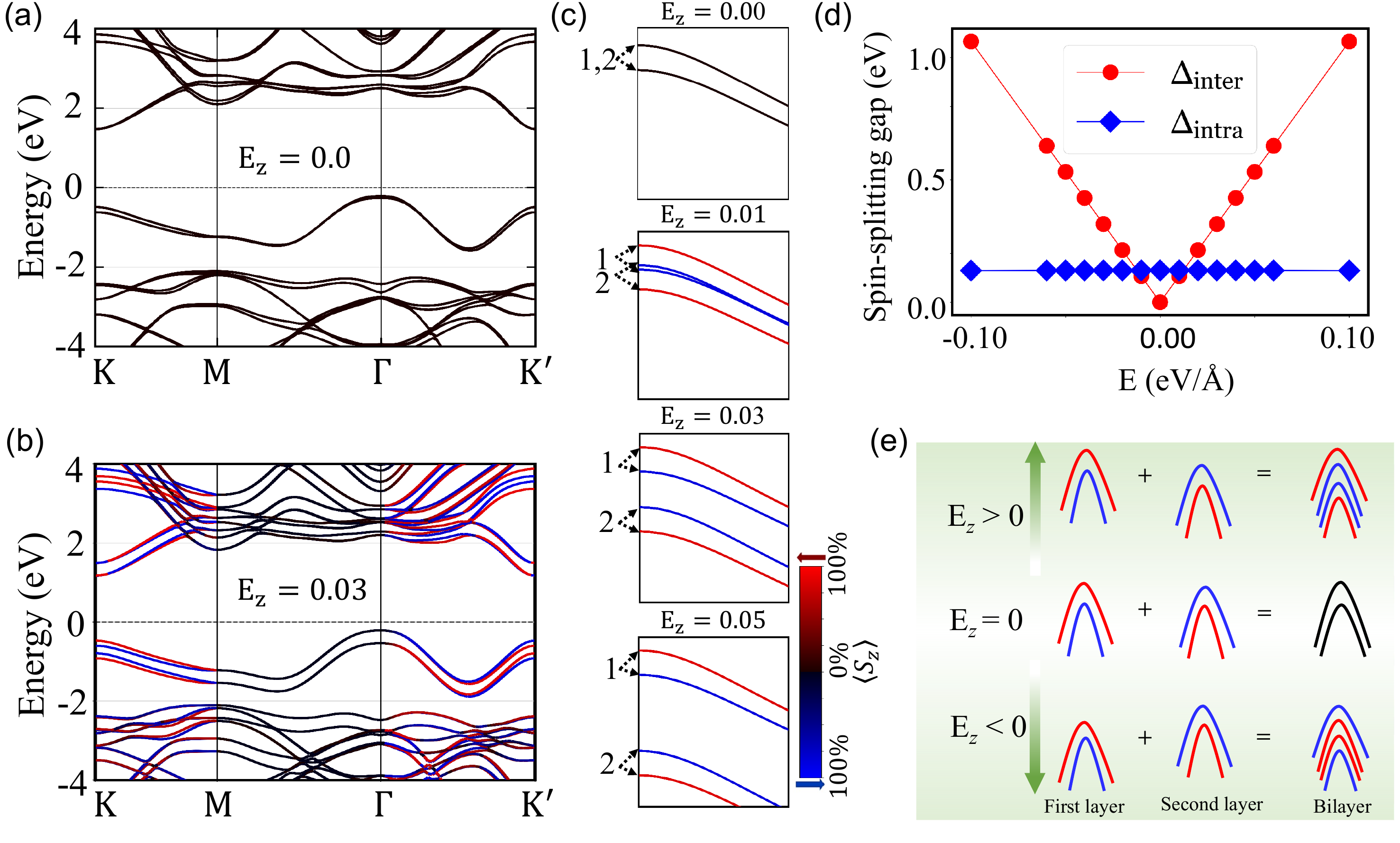}
\caption{(a) Band structure of bilayer MoSi$_2$N$_4$ in the absence of external electric field (E$_z$ = 0). (b) Same as (a) but for $E_z = 0.03$ eV/{\AA}. Spin-splitting in the band structure is evident. (c) Evolution of the top four valence bands around the $K$ point with varying external electric field strength. Color scale gives the degree (in percent) of spin-polarization of the bands. Markings 1 and 2 identify the doublets associated with the first and second layers of the bilayer. (d) Degree of spin-splitting at the $K$ point as a function of $E_z$. Blue (red) markers show the intra- (inter)-layer $\Delta_{intra}$ ($\Delta_{inter}$) components of the spin-splitting. (e) A schematic of the electric-field effect on the bilayer band structure.} \label{fig:BilayerElecSpin}
\end{figure}

{\it Tuning spin-structure of bilayer MoSi$_2$N$_4$ via an external electric field.}
Figure \ref{fig:BilayerElecSpin}(a) shows the band structure of bilayer MoSi$_2$N$_4$. Similar to the monolayer case, the bilayer is an indirect bandgap semiconductor with the VBM and CBM edges located at the $\Gamma$ and $K/K'$ points, respectively. However, in contrast to the monolayer, the inversion symmetry is now restored and, as a result, all bands become twofold spin-degenerate. A small splitting at the $\Gamma$ point is driven by the interlayer interactions between the two MoSi$_2$N$_4$ layers, whereas the splitting at the $K/K'$ points is due to the SOC. The inversion symmetry of the bilayer, however, can be broken by an out-of-plane external electric field $E_z$, which lifts the spin-degeneracy at the non-TRIM $K/K'$ points, allowing the manipulation of spin-split states at the top of the valence bands.

Figure~\ref{fig:BilayerElecSpin}(b) shows the spin-resolved bilayer band structure for $E_z=0.03$ eV/\AA. The spin-split states are now seen to be resolved at the $K$ and $K'$ points with opposite spin-polarizations for the top bands. There are four spin-polarized valence bands near the Fermi level, two of which originate from the first layer whereas the other two come from the second layer of the bilayer. Evolution of these four bands with $E_z$ is shown in Fig.~\ref{fig:BilayerElecSpin}(c). To quantify the spin-splitting, we introduce the quantities $\Delta_{intra}$ and $\Delta_{inter}$. Here, $\Delta_{intra}$ is defined as the energy difference between first (second) layer spin-up and first (second) layer spin-down states, while $\Delta_{inter}$ is the energy difference between the first-layer spin-up and second-layer spin-down states. $\Delta_{intra}$ thus captures the effect of the SOC on spin-splitting, whereas $\Delta_{inter}$ codes the effect of the potential difference between the two layers caused by the external field. When $E_z=0.01$ eV/\AA, the spin-split doublet from the second layer lies at an energy that is slightly lower than that for the first-layer doublet, so that  $\Delta_{inter} $ is smaller than $\Delta_{intra}$. The two topmost valence states are thus composed of states belonging to two different layers of the bilayer. When $E_z$ exceeds a critical value, $\Delta_{inter} $ becomes larger than $\Delta_{intra}$ and the two topmost valence states arise from the same layer. $\Delta_{intra}$ and $\Delta_{inter}$ are shown as a function of $E_z$ in Fig.~\ref{fig:BilayerElecSpin}(d). $\Delta_{inter}$ varies linearly with $E_z$ while $\Delta_{intra}$ shows negligible field dependence. A crossover between  $\Delta_{intra}$ and  $\Delta_{inter}$ is observed around $E_z=0.012$ eV/\AA. Notably, the spin polarization of the topmost valence states at the $K/K'$ points remains nearly 100\% in the presence $E_z$. 

\begin{figure}[t] % Bulk band structure of MoSi$_2$N$_4$
\centering  
\includegraphics[width=0.48\textwidth]{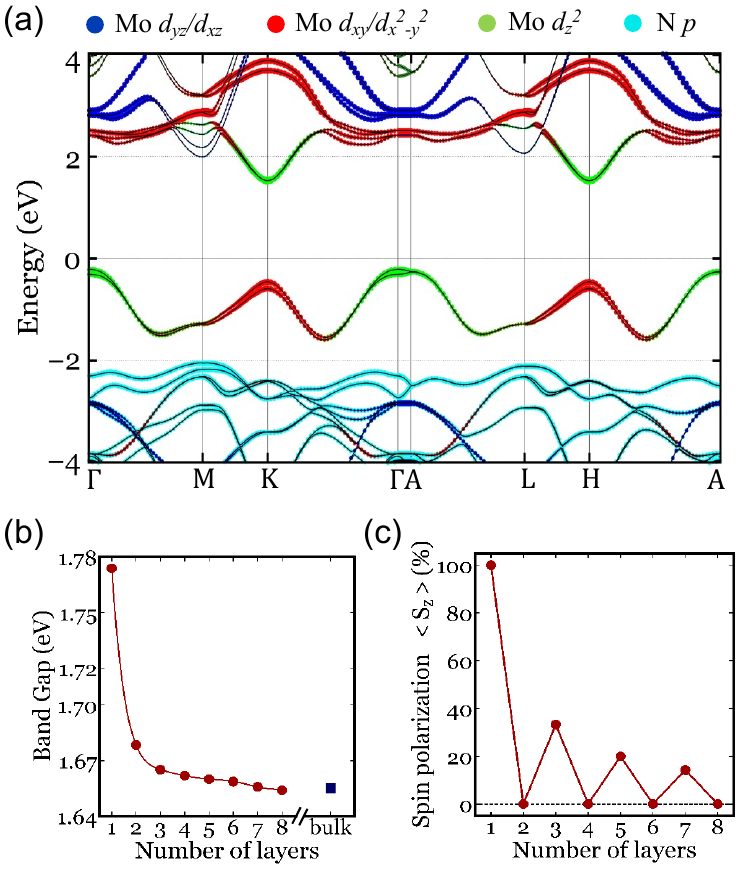}
\caption{(a) Orbitally-resolved band structure of bulk MoSi$_2$N$_4$ in the bulk hexagonal Brillouin zone. (b) Bandgap and (c) average spin-polarization of the top layer as a function of the number of layers.} \label{fig:Bulk}
\end{figure}

We find that the applied electric field changes the splitting ($\Delta_{inter}$) between the states coming from different layers in the bilayer. In contrast, as we would expect, the effect of the field on the spin-splitting as well as the degree of spin-polarization of the states coming from the same layer is negligible. Sign of the spin-polarization of states at $K/K'$ points is electric-field-direction dependent. Evolution of the states at the $K$ point under positive and negative field directions is shown schematically in Fig~\ref{fig:BilayerElecSpin}(e). These results provide a clear pathway for manipulating the spin states in bilayer MoSi$_2$N$_4$. Electric-field-dependent evolution of the bilayer states for all the MSi$_2$Z$_4$ materials we investigated falls along the preceding lines. Notably, the values of the electric field required to manipulate the states here are much lower than in MoS$_2$~\cite{Chang2014}.

{\it Layer-dependent states and spin polarization.}
We now turn to discuss the evolution of the bandgap and spin-polarization of multilayer MoSi$_2$N$_4$. Figure~\ref{fig:Bulk}(a) shows the calculated  bulk band structure using our optimized lattice parameters (Table~\ref{T1:bulk}). It has an indirect bandgap of
\rsp{$~1.655$ eV (2.221 eV) within the GGA (HSE)}. The wave functions at the CBM edge at $K$ and the VBM edge at $\Gamma$ consist of Mo $d_{z^2}$ states similar to the monolayer and bilayer cases. The bands along the $\Gamma-A$ direction remain weakly dispersive as a result of weak interlayer coupling. However, the SOC-split states can be seen at the $K$ and $H$ points. Evolution of the bandgap as a function of the layer thickness is shown in Fig.~\ref{fig:Bulk}(b). The bandgap decreases slightly with increasing number of MoSi$_2$N$_4$ layers and converges to the bulk value for the eight-layer film. This insensitivity of the bandgap to layer thickness indicates that the weak van der Waal's coupling dominates the interlayer interactions in MoSi$_2$N$_4$. 

Figure~\ref{fig:Bulk}(c) shows the evolution of spin-polarization of valence state as a function of the number of layers. Since the films with an even number of MoSi$_2$N$_4$ layers are inversion symmetric, these films display zero spin-polarization. Spin-polarization in films with an odd number of layers varies as $~1/N$, where $N$ is the number of layers.

\begin{figure}[!t]
\centering    
\includegraphics[width=0.5\textwidth]{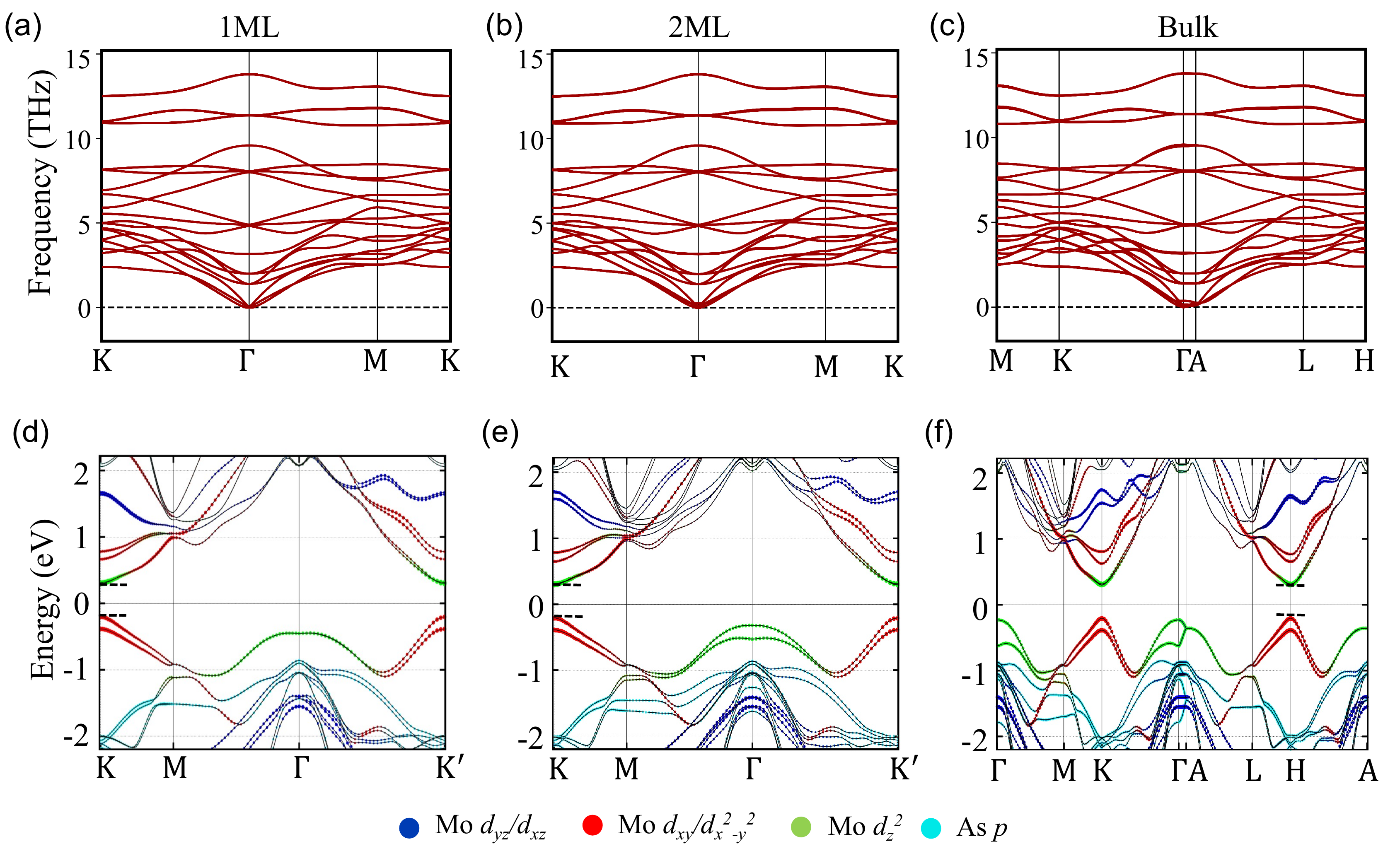}
\caption{Calculated phonon spectrum of (a) monolayer (1ML), (b) bilayer (2ML), and (c) bulk  MoSi$_2$As$_4$. Orbitally-decomposed band structure of (d) monolayer, (e) bilayer, and (f) bulk MoSi$_2$As$_4$. }\label{fig:MoSi2As4}
\end{figure}  

{\it Band structure of MSi$_2$Z$_4$ materials.}
We now discuss the dynamical stabilities and band structures of other MSi$_2$Z$_4$ thin films. Figures~\ref{fig:MoSi2As4}(a)-(c) show the phonon spectra of monolayer, bilayer, and bulk MoSi$_2$As$_4$. No imaginary branches in the BZ are found, indicating stability of these films. Band structures of MoSi$_2$As$_4$ films and bulk are presented in Figs.~\ref{fig:MoSi2As4}(d)-(f). In contrast to MoSi$_2$N$_4$, the monolayer MoSi$_2$As$_4$ is a direct bandgap semiconductor with a bandgap of 
\rsp{$~0.508$ eV (0.707 eV) within the GGA (HSE) } at the $K/K'$ point. The bandgap is found to remain direct as the thickness increases from monolayer to bulk \footnote{\rsp{The calculated HSE band structure with the optimized geometry obtained using the GGA and optPBE-vdW functional remains direct from monolayer to bulk MoSi$_2$As$_4$.}}. The interlayer coupling strength in MoSi$_2$As$_4$ is larger than in MoSi$_2$N$_4$, and the location of the direct bandgap changes from the $K$ to the $H$ point in going to the bulk limit [Fig.~\ref{fig:MoSi2As4}(f)]. MoSi$_2$As$_4$ monolayers also host nearly ~100\% spin-polarized states. 

The phonon spectra and orbitally-resolved band structures of WSi$_2$N$_4$ and WSi$_2$As$_4$ are presented in the SM~\cite{SPM}. These systems are also stable up to the bulk limit and support highly spin-polarized states similar to the cases of MoSi$_2$N$_4$ and MoSi$_2$As$_4$. However, the W atoms with their stronger SOC yield increased spin-splittings at the $K/K'$ points in these materials. 

{\it Conclusion.}
Using first-principles modeling, we have carried out a systematic thickness-dependent investigation of the dynamical stabilities and electronic and spin-polarization properties of the MSi$_2$Z$_4$ (M = Mo or W and Z = N or As) compounds. These materials are found to be dynamically stable from the monolayer to the bulk limit, indicating that multilayer films and bulk of such bottom-up synthesized 2D vdW materials should be possible to realize experimentally. Our analysis reveals that the monolayers host two nearly 100\% out-of-the-plane spin-polarized states at the $K$ points in the BZ with Zeeman-type spin splittings. The spin-polarization is reversed at the $K'$ points while the high degree of spin-polarization remains preserved. The spin-polarization of the states in the bilayers, which is zero due to the restoration of the inversion symmetry in the pristine bilayers, can be switched on and manipulated using an external electric field. MoSi$_2$N$_4$ and WSi$_2$N$_4$ exhibit a robust indirect bandgap from the monolayer to the bulk limit. In contrast, MoSi$_2$As$_4$ and WSi$_2$As$_4$ monolayers display a direct bandgap at the $K/K'$-point, which is preserved from the monolayer to the bulk. Our study provides insight into the bandgap, spin-polarization, and spin-valley locking of electronic states in MSi$_2$Z$_4$ materials class, and indicates that these materials could provide a viable materials platform as an alternative to the MoS$_2$ materials that are currently in common use for spintronics, valleytronics and optoelectronics applications.

\section*{Acknowledgments}

We thank Tomasz Dietl for valuable discussions. The work is supported by the Foundation for Polish Science through the international research agendas program co-financed by the European union within the smart growth operational program. We acknowledge the access to the computing facilities of the Interdisciplinary Center of Modeling at the University of Warsaw, Grant Nos. G75-10, GB84-0 and GB84-7. R. I.  and C. A. acknowledge support from Narodowe Centrum Nauki (NCN, National Science Centre, Poland) Project No.2020/37/N/ST3/02338. We acknowledge IIT Kanpur, Science Engineering and Research Board (SERB) and the Department of Science and Technology (DST) for financial support. The work at Northeastern University was supported by the Air Force Office of Scientific Research under award number FA9550-20-1-0322, and benefited from the computational resources of Northeastern University's Advanced Scientific Computation Center (ASCC) and the Discovery Cluster. The work at TIFR Mumbai is supported by the Department of Atomic Energy of the Government of India under project number 12-R$\&$D-TFR-5.10-0100.

\bibliography{MoSi2N4}
\end{document}